\documentclass[aps,prd,a4paper,reprint,amsmath,amssymb,superscriptaddress,nofootinbib]{revtex4-2}

\usepackage{hyperref}
\usepackage{graphicx}
\usepackage{cleveref}
\usepackage{booktabs}
\usepackage{xspace}

\usepackage{array}
\newcommand{\PreserveBackslash}[1]{\let\temp=\\#1\let\\=\temp}
\newcolumntype{C}[1]{>{\PreserveBackslash\centering}p{#1}}
\newcolumntype{R}[1]{>{\PreserveBackslash\raggedleft}p{#1}}
\newcolumntype{L}[1]{>{\PreserveBackslash\raggedright}p{#1}}

\usepackage[T1]{fontenc}
\usepackage[utf8]{inputenc}
\usepackage{lmodern}
\usepackage{microtype}

\makeatletter
\def\NAT@spacechar{\,}
\makeatother

\usepackage{xcolor}
\definecolor{theme}{rgb}{0.26,0.41,0.88}
\hypersetup{
    colorlinks=true,
    allcolors=theme,
}

\usepackage{bm}
\let\vec\bm

\newcommand{\Ccite}{Ref.~\cite}
\newcommand{\ccite}{ref.~\cite}

\newcommand{\mHuSq}{\ensuremath{m^2_{\hspace{-0.1em}\scriptscriptstyle H_{\hspace{-0.1em}\scriptscriptstyle u}}\hspace{-0.1em}}}

\newcommand{\mHdSq}{\ensuremath{m^2_{\hspace{-0.1em}\scriptscriptstyle H_{\hspace{-0.1em}\scriptscriptstyle d}}\hspace{-0.1em}}}

\let\GreekDelta\Delta
\renewcommand{\Delta}{\triangle}
\newcommand{\DKL}{\ensuremath{\mathcal{D}}}
\newcommand{\BG}{\ensuremath{\GreekDelta}}
\newcommand{\namebg}{traditional FT\xspace}
\newcommand{\Namebg}{Traditional FT\xspace}

\newcommand{\given}{\,|\,}
\newcommand{\like}{\mathcal{L}}
\newcommand{\ev}{\mathcal{Z}}
\newcommand{\nleft}{\mathopen{}\mathclose\bgroup\left}
\newcommand{\nright}{\aftergroup\egroup\right}

\newcommand{\moment}[1]{\mathbb{E}\big[\vphantom{\mathbb{E}}#1\big]}

\newcommand{\gev}{\,\text{GeV}}
\newcommand{\tev}{\,\text{TeV}}
\newcommand{\nats}{\,\text{nats}}
\renewcommand{\nat}{\,\text{nat}}
\newcommand{\bits}{\,\text{bits}}

\newcommand{\code}[1]{\texttt{#1}}

\newcommand{\intd}{\,\text{d}}
\DeclareMathOperator{\sign}{sign}
\newcommand{\ifrac}[2]{#1/#2}
\newcommand{\vtheta}{\vec\theta}
\newcommand{\eqntext}[1]{\text{\small #1}}

\makeatletter
\newlength{\negph@wd}
\DeclareRobustCommand{\negphantom}[1]{%
  \ifmmode
    \mathpalette\negph@math{#1}%
  \else
    \negph@do{#1}%
  \fi
}
\newcommand{\negph@math}[2]{\negph@do{$\m@th#1#2$}}
\newcommand{\negph@do}[1]{%
  \settowidth{\negph@wd}{#1}%
  \hspace*{-\negph@wd}%
}
\makeatother

\begin{document}

\title{Precise interpretations of traditional fine-tuning measures}

\author{Andrew Fowlie}
\affiliation{Department of Physics, School of Mathematics and Physics, Xi'an Jiaotong-Liverpool University, Suzhou, 215123, China}
\email{andrew.fowlie@xjtlu.edu.cn}
 
\author{Gonzalo Herrera}
\affiliation{Center for Neutrino Physics, Department of Physics,\\ Virginia Tech, Blacksburg, VA 24061, USA}
\email{gonzaloherrera@vt.edu}

\begin{abstract}
We uncover two precise interpretations of traditional electroweak fine-tuning (FT) measures that were historically missed. \emph{(i) a statistical interpretation}: the \namebg measure shows the change in plausibility of a model in which a parameter was exchanged for the $Z$ boson mass relative to an untuned model in light of the $Z$ boson mass measurement. \emph{(ii) an information-theoretic interpretation:} the \namebg measure shows the exponential of the extra information, measured in nats, relative to an untuned model that you must supply about a parameter in order to fit the $Z$ mass. We derive the mathematical results underlying these interpretations, and explain them using examples from weak scale supersymmetry. These new interpretations allow us to rigorously define FT in particle physics and beyond; shed fresh light on the status of extensions to the Standard Model; and lastly, allow us to precisely reinterpret historical and recent studies using \namebg measures.
\end{abstract}

\maketitle

\section{Introduction}

Although the concepts of fine-tuning and naturalness are under pressure~\cite{Richter:2006um,Dine:2015xga,Hossenfelder:2018ikr}, they remain popular among physicists~\cite{tHooft:1979rat,ef2165c8-7176-35d7-a191-db6afd794cac,Feng:2013pwa,Giudice:2008bi,Craig:2022eqo,Wells:2018yyb}. Efforts to define and measure fine-tuning began in the context of supersymmetric models in the 1980s. \Ccite{Ellis:1986yg,Barbieri:1987fn} introduced a measure of fine-tuning of the parameter $a_i$ to fit the $Z$ mass,
\begin{equation}\label{eq:bg}
   \BG_i = \left|\frac{\text{d} \ln M_Z}{\text{d} \ln a_i}\right| = \left|\frac{a_i}{M_Z} \frac{\text{d} M_Z}{\text{d} a_i}\right|.
\end{equation}
We refer to this as the traditional fine-tuning (FT) measure.\footnote{\Cref{eq:bg} may be more commonly known as the Barbieri-Giudice measure or as the Barbieri-Giudice-Ellis-Nanopoulos measure.} The \namebg measure shows sensitivity in the parameter $a_i$. To aggregate \namebg measures for more than one  parameter, \ccite{Ellis:1986yg,Barbieri:1987fn} suggested maximizing across them,
\begin{equation}\label{eq:max_bg}
   \BG = \max_i \BG_i = \max_i \left|\frac{a_i}{M_Z} \frac{\text{d} M_Z}{\text{d} a_i}\right|.
\end{equation}
There are, however, other possibilities, e.g., adding \namebg measures in quadrature~\cite{Casas:2004gh}. By these measures, fine tuning in supersymmetric models became severe after results from LEP~\cite{Anderson:1994tr,Chankowski:1997zh,Barbieri:1998uv,Chankowski:1998xv,Kane:1998im} and even worse after results from LHC runs I and II~\cite{Strumia:2011dv,Baer:2012mv,Arvanitaki:2013yja,Baer:2014ica,vanBeekveld:2019tqp}, 
including the discovery of the Higgs boson~\cite{Strumia:2011dv,Baer:2012mv,Arvanitaki:2013yja,Baer:2014ica,vanBeekveld:2019tqp}. 
The severity of fine tuning led to doubts about weak scale supersymmetry; however, the measure of fine tuning appears somewhat arbitrary and lacking in precise meaning. Efforts to connect fine tuning with a probability of cancellations~\cite{Anderson:1994dz,Ciafaloni:1996zh,Giusti:1998gz,Strumia:1999fr,Allanach:2006jc,Athron:2007ry} ultimately led to an interpretation of fine tuning in Bayesian inference~\cite{Allanach:2007qk,Cabrera:2008tj,
Cabrera:2010dh,Balazs:2012vjy,Fichet:2012sn,Fowlie:2014faa,Fowlie:2014xha,
Fowlie:2015uga,Clarke:2016jzm,Fowlie:2016jlx,Athron:2017fxj,Fundira:2017vip}. Indeed, \namebg measures were shown to appear as factors in integrands in intermediate stages of Bayesian inference. A precise interpretation of the measures themselves, however, remained lacking (see \cref{app:history} for further discussion). 

We recently proposed the Bayes factor (BF) surface~\cite{Fowlie:2024dgj} as a new way to understand the impact of experimental measurements on models of new physics.\footnote{See~\ccite{gram2020,Johnson_2023,Wagenmakers2020,Pawel2023,NANOGrav:2023hvm} for recent related works in other contexts.} Considering the measurement of the $Z$ mass, the BF surface reveals a precise statistical interpretation of traditional fine-tuning measures. This provides justification for traditional fine-tuning arguments about the weak scale.

We have, furthermore, discussed in recent years applications of information theory in interpreting experimental searches of new physics, e.g., \ccite{Fowlie:2017ufs, Fowlie:2018svr, Herrera:2024zrk}. We show that \namebg measures show exactly the exponential of the extra information, measured in nats, required to predict the $Z$ mass. This is a fresh insight into the meaning of \namebg measures.

\begin{figure*}[t]
    \centering
    \includegraphics{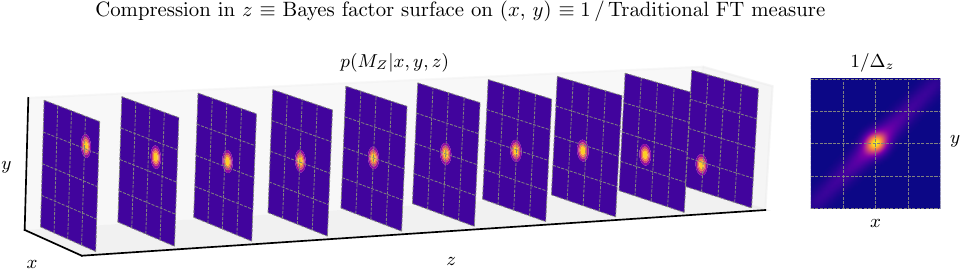}
    \caption{The likelihood of the $Z$ mass measurement in a three-dimensional toy model. The compression required in the $z$ parameter relative to an untuned model equals both the Bayes factor surface relative to an untuned model and the reciprocal of the Barbieri-Giudice fine-tuning measure with respect to $z$.}
    \label{fig:cartoon}
\end{figure*}

\section{Interpreting \texorpdfstring{$Z$}{Z} mass measurement}

\subsection{Bayes factor surface}

The Bayes factor (BF) surface~\cite{Fowlie:2024dgj} shows the change in plausibility of a model as a function of that model's parameters relative to a reference model. For the evidence in favor of the reference model, we may write this as
\begin{equation}
    B(\vtheta) = \frac{\ev_0}{\ev_1(\vtheta)},
\end{equation}
where $\ev_1(\vtheta)$ and $\ev_0$ are the evidences for the model and reference model, respectively. The evidences themselves are likelihoods averaged over a choice of prior~\cite{Jeffreys:1939xee,kass1995bayes};
\begin{equation}
    \ev = \int \like(\vtheta) \pi(\vtheta) \intd \theta,
\end{equation}
for likelihood $\like$ and prior $\pi$.

First, consider a trivial toy model with a single parameter $\phi$, that predicts the $Z$ mass as
\begin{equation}
    M_Z^2 = \phi^2.
\end{equation}
We refer to this as an \emph{untuned model}. The narrow Gaussian for the combined $Z$ mass measurement, $\hat M_Z = 91.1876\pm0.0021\gev$~\cite{Workman:2022ynf}, may be approximated by a Dirac function,
\begin{equation}\label{eq:dirac_mz}
    \like(M_Z) = \delta(M_Z - \hat M_Z).
\end{equation}
Thus the evidence for this untuned model may be written,
\begin{equation}\label{eq:z_0}
    \ev_0 = \int \like(M_Z) \, \pi(\phi) \intd \phi = \pi(\hat\phi) = \pi(M_Z),
\end{equation}
where $\hat\phi = M_Z$ such that $M_Z(\hat \phi) = \hat M_Z$.

Second, consider a \emph{complicated model} with parameters $\vec\Phi$ that predicts the $Z$ mass as
\begin{equation}
    M_Z^2 = M_Z^2(\vec\Phi).
\end{equation}
This model could be, for example, weak-scale supersymmetry. To predict the correct $Z$ mass, we assume that the parameter $\phi = \Phi_i$ was exchanged for the $Z$ mass. We denote the remaining parameters by $\vtheta$. For this complicated model,
\begin{equation}\label{eq:z_i}
    \begin{split}
        \ev_i(\vtheta) = \int \like(M_Z) \, &\pi(\phi \given \vtheta) \intd \phi\\
        & = \left|\frac{\text{d} M_Z}{\text{d} \phi} \, \pi(\phi \given \vtheta) \right|_{\phi \,=\, \hat\phi},        
    \end{split}
\end{equation}
where $\hat \phi$ is a function of $\vtheta$ such that $M_Z(\vtheta, \hat \phi) = \hat M_Z$. 
We consider the impact of approximation \cref{eq:bf_mz} in \cref{app:gaussian}. We show that when we treat the likelihood as a Gaussian $\hat M_Z \pm \sigma$, corrections to the evidence depend only on second-order variations in the prior prediction for the $Z$ mass around $\hat M_Z \pm \sigma$.

We now consider the BF between the untuned and complicated model as a function of $\vtheta$ --- this is the BF surface.  Using \cref{eq:z_0,eq:z_i}, we find
\begin{equation}\label{eq:bf_pi_general}
   B_i(\vtheta) = \frac{\ev_0}{\ev_i(\vtheta)} = \left|\frac{\text{d} M_Z}{\text{d} \phi} \frac{\pi(M_Z)} {\pi(\phi \given \vtheta)}\right|_{\phi \,=\,\hat\phi}.
\end{equation}
Our notation indicates that the $\phi = \Phi_i$ parameter was marginalized.
\Cref{eq:bf_pi_general} depends on choices of prior for the model parameters. 
The scale-invariant density, $p(x) \propto 1/x$, weights every order of magnitude equally~\cite{4082152,Hartigan1964,Consonni2018}. Consider an identical scale-invariant prior for $\phi$ in the untuned and complicated model, 
\begin{equation}
    \pi(\phi) = \pi(\phi \given \vtheta) = \begin{cases}
                         \frac{1}{\ln(b/a)
                         } \frac{1}{|\phi|} & a \le \phi \le b \\
                        0 & \text{otherwise}
                    \end{cases},\label{eq:log_prior}
\end{equation}
over the range $a$ to $b$. With this choice, we obtain,
\begin{equation}\label{eq:bf_mz}
   B_i(\vtheta) =  \left|\frac{\phi}{M_Z} \frac{\text{d} M_Z}{\text{d} \phi}\right|_{\phi \,=\, \hat\phi},
\end{equation}
where the dependence on $a$ and $b$ canceled.  This prior is not compatible with parameter values of zero, e.g.~the supersymmetric parameter $A_0 = 0$, as it would be an improper prior.

\subsection{Information}

The Kullback-Leibler (KL) divergence~\cite{Kullback:1951zyt} between the prior and the posterior for a model parameter $\phi = \Phi_i$ may be written as,
\begin{equation}\label{eq:dkl}
    \DKL_i \equiv \int p(\phi \given \vtheta, \hat M_Z) \ln \nleft(\frac{p(\phi \given \vtheta,\hat M_Z)}{\pi(\phi \given \vtheta)}\nright) \intd \phi.  
\end{equation}
The KL divergence can be interpreted as a measure of information learned about $\phi$ from the measurement of the $Z$ mass or compression from the prior to the posterior~\cite{mackay2003information}. Consider, for example, a one-dimensional model with a unit uniform prior for parameter~$\phi$. Suppose that we measured $\phi = 1/2 \pm 1/128$. We would find that $\DKL \simeq \ln 128$ --- we learned $\ln 128 \nats = 7\bits$ of information about $\phi$ and compressed by $e^{\DKL} \simeq 128$.\footnote{With a natural logarithm in the KL divergence, as in \cref{eq:dkl}, information is measured in nats; with a base 2 logarithm, it is measured in bits. They are connected by $1\nat = 1/\ln2 \bits$.} 

Using \cref{eq:dkl} and recognizing that by Bayes' theorem, 
$\ifrac{p(\phi \given \vtheta, \hat M_Z)}{\pi(\phi \given \vtheta)} = \ifrac{\like(\phi, \vtheta)}{\ev}$, we may express the evidence as~\cite{Hergt:2021qlh},
\begin{equation}\label{eq:z_decomposition}
    \ln \ev_i = \langle \ln \like\rangle - \DKL_i.
\end{equation} 
where $\langle \cdot \rangle$ indicates a posterior mean. Expressing \cref{eq:z_decomposition} in words reveals a Bayesian form of Occam's razor~\cite{Hergt:2021qlh,Rosenkrantz1977,37c684a3-9d55-37ab-ba95-6e4da1d0a51b,1991BAAS...23.1259J,Henderson_Goodman_Tenenbaum_Woodward_2010,4018651d-dd41-3908-8c4f-09d2118c5612,10.1093/oxfordhb/9780199957996.013.14,mackay2003information,Blanchard2017,McFadden2023},
\begin{equation}
\begin{split}
   \eqntext{Weight}&\eqntext{ of evidence} =\\
   &\eqntext{Average goodness-of-fit} - \eqntext{fine-tuning cost},
\end{split}
\end{equation}
as models are penalized for fine tuning measured by the KL divergence. From \cref{eq:z_decomposition}, furthermore, the Bayes factor may be written,\footnote{To avoid ambiguity, we denote differences by $\Delta$ and the \namebg measure by $\BG$.}
\begin{equation}\label{eq:b_decomposition}
    \ln B = \Delta \ln \ev = \Delta \langle \ln \like\rangle - \Delta \DKL.
\end{equation} 
For the measurement of the $Z$ mass, we find that
\begin{equation}
    \ln B = -\Delta \DKL.
\end{equation}
This happens because, regardless of the model, a posteriori it must predict that $M_Z = \hat M_Z$ because of the Dirac function in \cref{eq:dirac_mz}. This means that the goodness-of-fit contributions to the Bayes factor in \cref{eq:b_decomposition} cancel.
In \cref{app:gaussian}, we checked average goodness-of-fit for a Gaussian likelihood, $\hat M_Z \pm \sigma$. The cancellation holds up to corrections that depend only on second-order variations in the prior prediction for the $Z$ mass around $\hat M_Z \pm \sigma$.

\section{Interpreting fine tuning}

\begin{table*}[t]
    \centering
    \begin{tabular}{L{2cm}L{3.75cm}L{1.5cm}L{2cm}L{1.5cm}L{3.5cm}}
    \toprule
    \multicolumn{2}{c}{Jeffreys~\cite{Jeffreys:1939xee}} &  \multicolumn{2}{c}{Lee \& Wagenmakers~\cite{lee2014bayesian}} & \multicolumn{2}{c}{Kass \& Raftery~\cite{kass1995bayes}} \\
    \cmidrule(r){1-2}  \cmidrule(r){3-4}  \cmidrule(r){5-6}
    $1$ to $10^{1/2}$ & Barely worth mentioning & 1 to 3 & Anecdotal & 1 to 3 & Barely worth mentioning\\
    $10^{1/2}$  to $10$ & Substantial & 3 to 10 & Moderate & 3 to 20 & Positive\\
    $10$ to $10^{3/2}$ & Strong & 10 to 30 & Strong & 20 to 150 & Strong\\
    $10^{3/2}$ to $100$ & Very strong & 30 to 100 & Very strong & $>150$ & Very strong \\
    $>100$ & Decisive & $> 100$ & Extreme\\
    \bottomrule
    \end{tabular}
    \caption{The Jeffreys~\cite{Jeffreys:1939xee}, Lee \& Wagenmakers~\cite{lee2014bayesian}, and Kass \& Raftery~\cite{kass1995bayes} scales for interpreting Bayes factors. Independently, the thresholds $10$~\cite{Barbieri:1987fn} and $30$~\cite{Baer:2015rja} were proposed for fine-tuning measures.}
    \label{tab:scales}
\end{table*}

\subsection{Equivalence}

Using \cref{eq:bf_mz,eq:b_decomposition,eq:bg}, we thus find
\begin{equation}\label{eq:bfs_dkl_bg}
 B_i = e^{\Delta \DKL}  = \BG_i  
\end{equation}
or in words, using the parameter $i$ to fix the $Z$ mass,
\begin{equation}\label{eq:bfs_dkl_bg_words}
\begin{split}
      \eqntext{Bayes factor} &= \eqntext{Relative compression} \\
                             &= \eqntext{\Namebg}  
\end{split}
\end{equation}
relative to an untuned model that predicts $M_Z^2 = \phi^2$. This required us to choose identical scale-invariant priors for the parameters exchanged for the $Z$ mass and approximate the narrow Gaussian measurement of the $Z$ mass by a Dirac function, though see \cref{app:requirements} where we justify these requirements. In particular, in \cref{app:gaussian} we checked the Dirac function approximation to a Gaussian likelihood. We found that corrections depended only on second-order variations of the prior prediction for the $Z$ mass around the peak of the Gaussian measurement.

\Cref{eq:bfs_dkl_bg,eq:bfs_dkl_bg_words} are the main results of this work. To the best of our knowledge, this clear connection between fine-tuning, Bayesian statistics and information theory had not been pointed out earlier in the literature.

\subsection{Interpretation}

\Cref{eq:bfs_dkl_bg} provides two precise interpretations of the \namebg measure,
\begin{itemize}
    \item \emph{Statistical} --- the \namebg measure shows the Bayes factor surface versus an untuned model. That is, it measures the change in plausibility of a model relative to an untuned model in light of the $Z$ mass measurement.
   
    \item \emph{Information-theoretic} --- the \namebg measure shows the compression versus an untuned model. That is, it measures the exponential of the extra information, measured in nats, relative to an untuned model that you must supply about a parameter in order to fit the $Z$ mass.\footnote{Similarly, \ccite{Dermisek:2016zvl} proposed counting the number of digits to which a parameter must be specified to predict the correct $Z$ mass.}
\end{itemize}
For illustration, in \cref{fig:cartoon} we show these interpretations in a three-dimensional toy model. The tuning and compression required in the $z$-direction changes across the ($x$, $y$) plane. The \namebg measure on the ($x$, $y$) plane shows the required compression.

Bayes factors themselves may be interpreted by ascribing qualitative meanings to them. \Cref{tab:scales} shows scales for interpreting a Bayes factor~\cite{Jeffreys:1939xee,lee2014bayesian,kass1995bayes}. Jeffreys~\cite{Jeffreys:1939xee} and Lee \& Wagenmakers~\cite{lee2014bayesian} take $10^{3/2} \approx 30$ and $30$, respectively, as a threshold for \emph{very strong} evidence, and $10$ as a threshold of \emph{moderate} evidences. The thresholds $10$ and $30$ were chosen as thresholds for fine-tuning measures in~\ccite{Barbieri:1987fn} and~\cite{Baer:2015rja}, respectively.

Lastly, \namebg measures are frequently reported as a percentage tuning, $100\% / \BG_i$, such that, e.g., $\BG_i = 100$ represents a $1\%$ tuning. In fact, the percentage tuning approximately equals the posterior probability of the complicated model,
\begin{equation}\label{eq:posterior}
    P_1 = \frac{\pi_1 \ev_1}{\pi_0 \ev_0 + \pi_1 \ev_1} = \frac{1}{\pi_0 / \pi_1 \BG_i + 1}\simeq 1 / \BG_i,
\end{equation}
where $\pi_0$ and $\pi_1$ are the prior probabilities of the untuned and complicated models, respectively. The approximation holds whenever $\BG_i \gg 1$, and the untuned and complicated models are the only models under consideration and are equally plausible a priori. In these cases, a tuning of e.g., $\BG_i = 100$ means that the complicated model was reduced to $1\%$ plausibility by the $Z$ mass measurement.

\subsection{Choice of parameter}

To apply the \namebg measure in \cref{eq:bg}, one must choose a parameter with which to take a derivative, or a way in which to combine \namebg measures for all parameters, such as \cref{eq:max_bg}. Similarly, in the Bayes factor surface, one parameter must be marginalized to exchange for the $Z$ mass. This choice defines different Bayes factor surfaces. For example, in the context of the constrained minimal supersymmetric Standard Model (CMSSM) with parameters $m_{1/2}$, $m_0$, $A_0$, $\tan\beta$ and $\mu$, we could consider the surface for $m_{1/2}$, $m_0$, $A_0$ and $\tan\beta$ for $\mu$ marginalized, or the surface for $\mu$, $m_0$, $A_0$ and $\tan\beta$ for $m_{1/2}$ marginalized. 
To compute the evidence, however, one would marginalize all parameters.

On the other hand, by \cref{eq:bfs_dkl_bg}, the maximum \namebg measure may be interpreted as 
\begin{itemize}
    \item \emph{Statistical} --- the Bayes factor against the worst model that fits the $Z$ mass by exchanging one parameter.
    
    \item \emph{Information-theoretic} --- the exponential of the maximum extra information, measured in nats, that must be specified to fit the $Z$ mass by exchanging one parameter.
\end{itemize}
The Bayes factor and extra information are both relative to an untuned model.

\section{Examples}

We now present examples from supersymmetric models --- the original context in which the \namebg measure was proposed. We use full one-loop and leading two-loop electroweak symmetry breaking conditions and two-loop renormalization group equations.\footnote{We used the \code{MssmSoftsusy::fineTune} method from \code{SOFTSUSY-4.1.20}~\cite{Allanach:2001kg}.} 

\subsection{Minimal Supersymmetric Standard Model}

The minimal supersymmetric Standard Model~(MSSM) predicts that at tree-level~\cite{Martin:1997ns}
\begin{equation}\label{eq:dirac_mz_mssm}
    \frac12 M_Z^2 = \frac{\mHdSq - \mHuSq \tan^2\beta}{\tan^2\beta - 1} - \mu^2,
\end{equation}
where $\mHuSq$ and $\mHdSq$ are soft-breaking Higgs mass parameters, $\tan\beta$ is the ratio of Higgs vacuum expectation values, and $\mu$ is a supersymmetry-preserving Higgs mass parameter.

We take a well-tempered neutralino benchmark in the MSSM from Snowmass 2013~\cite{Cahill-Rowley:2013gca}. In this scenario, the neutralino is a well-tempered admixture of bino-higgsino and plays the role of dark matter. The tuning for this benchmark with respect to the $\mu$-parameter, $\BG_\mu \simeq 20$, indicates that, if exchanging $\mu$ for the $Z$ mass, 
\begin{itemize}
    \item \emph{Statistical} --- This scenario is disfavored by a factor of about 20 relative to an untuned model.
    \item \emph{Information-theoretic} --- An extra $\ln20\nats \simeq 4\bits$ of information must be supplied about the $\mu$ parameter to fit the $Z$ mass relative to an untuned model.
\end{itemize}
The worst tuning, $\BG \gtrsim 1000$, was with respect to the gluino mass parameter, $M_3$. If this parameter were treated as unknown and marginalized in exchange for the $Z$ mass, the scenario would be disfavored by more than 1000 and more than an extra $10\bits$ of information about $M_3$ would be required to fit the $Z$ mass.

\subsection{Constrained MSSM}

We consider the constrained MSSM (CMSSM;~\cite{Kane:1993td,Chamseddine:1982jx}). The MSSM soft-breaking terms are unified at the unification sale, leaving a universal scalar soft-breaking mass $m_0$, a universal gaugino soft-breaking mass $m_{1/2}$, and a universal trilinear term $A_0$. In \cref{fig:bg_measure}, we show the $\tan\beta=50$, $A_0=0$ and $\sign\mu=1$ slice of the ($m_0$, $m_{1/2}$) plane. The \namebg measure for the $\mu$ parameter, $\BG_\mu$, increases from around 1 to over 5000 as $m_{1/2}$ increases. Besides the informal interpretation that the \namebg measure represents undesirable and unnatural fine tuning in the $\mu$ parameter, there are two precise and exact interpretations:
\begin{itemize}
    \item \emph{Statistical} --- The \namebg measure shows the decrease in plausibility in this model in which $\mu$ was unknown relative to an untuned model.
    \item \emph{Information-theoretic} --- The \namebg measures the extra information that must be supplied about the $\mu$ parameter to fit the $Z$ mass relative to an untuned model.
\end{itemize}
For example, relative to an untuned model, models with $\BG_\mu > 300$ decrease in plausibility by more than a factor 300 in light of the $Z$ mass, and at least an extra $\ln 300\nats \simeq 6 \nats \simeq 8 \bits$ of information must be supplied about the $\mu$ parameter to fit the $Z$ mass.

Heavier gaugino masses result in fine-tuning as gaugino masses contribute radiatively to terms in \cref{eq:dirac_mz_mssm}. \Cref{fig:bg_measure} shows, however, a narrow strip of parameter points fine-tuned by $\BG_\mu \lesssim 10$ that extends to multi-TeV --- this is the focus point region~\cite{Feng:1999zg,Feng:1999mn,Feng:2000gh,Feng:2011aa}. As we consider $\tan\beta = 50$, \cref{eq:dirac_mz_mssm} may be approximated by
\begin{equation}
    \frac12 M_Z^2 \approx  - \mHuSq  - \mu^2.
\end{equation}
In the focus point region, the renormalization group equations (RGEs) focus the soft-breaking supersymmetric masses at the weak scale. That is, at the weak scale the soft-breaking Higgs mass $\mHuSq \sim M_Z^2$ regardless of the ultraviolet boundary condition for $m_0$. 
This focusing means that we do not need to fine tune cancellations between $\mHuSq$ and $\mu^2$.

\begin{figure}[t]
    \centering
    \includegraphics[width=\linewidth]{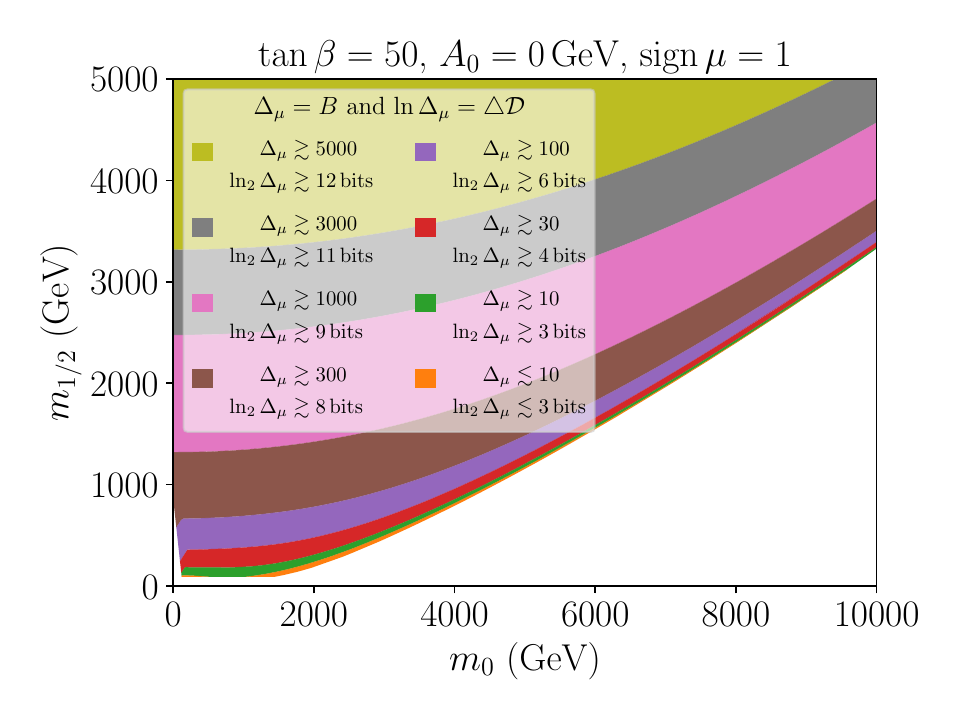}
    \caption{The \namebg measure with respect to $\mu$ for the $\tan\beta=50$, $A_0=0$ and $\sign\mu=1$ slice of the ($m_0$, $m_{1/2}$) plane of the CMSSM. This is equivalent to the BF surface versus an untuned model, $B$, and the extra information required about $\mu$, $\Delta\DKL$.}
    \label{fig:bg_measure}
\end{figure}

\subsection{Combining fine-tuning and the Higgs mass measurement}

The \namebg measure results from applying the $Z$ mass measurement to a model in a Bayesian framework. This framework tells us how to combine it with other measurements. For example, we now consider the Higgs mass measurement, $\hat m_h = 125.25\pm0.17\gev$~\cite{Workman:2022ynf}. This Higgs mass can be realized in the CMSSM through loop corrections from heavy sparticles, at the cost of fine tuning. The statistical interpretation of the \namebg measure sheds light on this interplay. 

\begin{figure}[t]
    \centering
    \includegraphics[width=\linewidth]{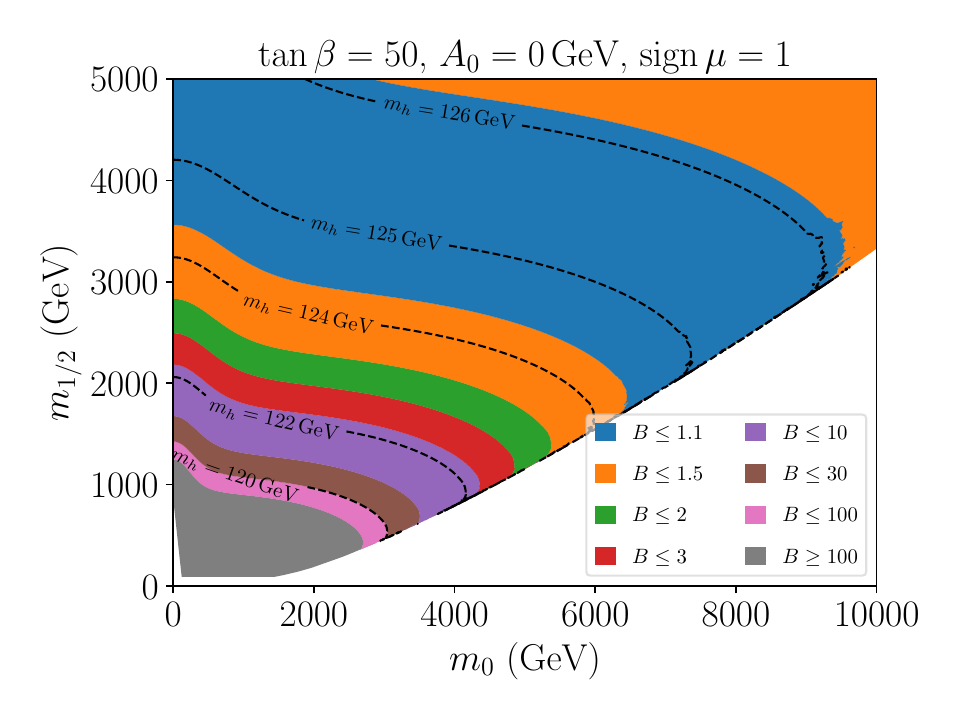}
    \caption{Bayes factor surface for the Higgs mass measurement, $\hat m_h = 125.25\pm0.17\gev$~\cite{Workman:2022ynf}, for a slice of the CMSSM relative to a model that predicts $m_h = \hat m_h$.}
    \label{fig:higgs}
\end{figure}

We suppose that our toy \emph{untuned model} predicts that $m_h = \hat m_h$, and use a two-loop prediction from the CMSSM from \code{SOFTSUSY-4.1.20}~\cite{Allanach:2001kg}. We assume that both predictions are known to within an uncertainty of $2\gev$. For illustration, we show the BF surface for the Higgs mass measurement in \cref{fig:higgs}. Parameters with $m_{1/2} \gtrsim 3\tev$ and in the narrow focus point strip predict $m_h \simeq 125\gev$ and are just as good as the untuned model, thus $B \simeq 1$. 

We should, however, apply the $Z$ and Higgs mass measurements simultaneously. To combine them, we multiply the BF surface for the Higgs mass by the \namebg measure for the $Z$ mass,\footnote{We neglect common nuisance parameters, e.g., the mass of the top quark. If they were included, the BF surfaces could not be simply multiplied.} 
\begin{align}
    B_{hZ}(\vtheta) &= \frac{p(m_h, M_Z \given M_0)}{p(m_h, M_Z \given M_1, \vtheta)} \\
    &= \frac{p(m_h \given M_Z, M_0)}{p( m_h \given M_Z, M_1, \vtheta)}  \times \frac{p(M_Z \given M_0)}{p( M_Z \given M_1, \vtheta)} \\
    &= B_h(\vtheta) \times \BG(\vtheta) \vphantom{\frac{p(M_Z \given M_0)}{p( M_Z \given M_1, \vtheta)}}.
\end{align}
We retain the statistical interpretation, though lose the information-theoretic one as the BF surface for the Higgs mass contains both relative information and goodness-of-fit contributions.

We see in \cref{fig:higgs_and_bg} that the resulting BF surface favors the untuned model by more than $1000$, except in the focus point strip. For focus points, the untuned model could be favored by less than a factor of $10$. 
The focus point, however, occupies a narrow strip of the ($m_0$, $m_{1/2}$) plane. If we were to consider the model evidence by averaging over the entire plane weighted by a choice of prior density, the impact of the focus point could be negligible as it occupies a negligible prior volume.

\begin{figure}[t]
    \centering
    \includegraphics[width=\linewidth]{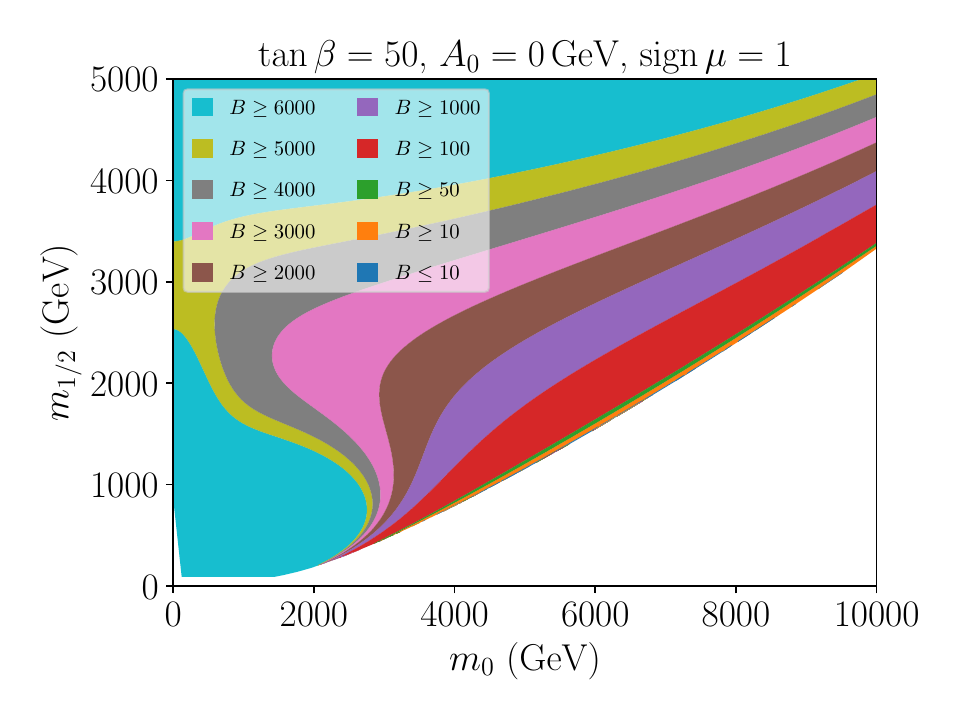}
    \caption{Bayes factor surface for the $Z$ and Higgs mass measurements for a slice of the CMSSM relative to an untuned model.}
    \label{fig:higgs_and_bg}
\end{figure}

\section{Conclusions}

Fine-tuning measures have played an important role in our assessment of theories of new physics since the 1980s. For example, doubts were raised about weak scale supersymmetric theories based on calculations of the fine-tuning prices of LEP in the 1990s, and more recently the LHC. For the first time, we provide precise interpretations of the traditional measure of fine tuning:
\begin{itemize}
    \item \emph{Statistical} --- the \namebg measure shows us the change in plausibility of a model in which one parameter was exchanged for the $Z$ mass relative to an untuned model in light of the $Z$ mass measurement.
   
    \item \emph{Information-theoretic} --- the \namebg measure shows the exponential of the extra information, measured in nats, relative to an untuned model that you must supply about a parameter in order to fit the $Z$ mass.
\end{itemize}
We used these interpretations to understand fine-tuning in the CMSSM parameter space. We found that the fine-tuning required to fit the $Z$ mass constitutes strong evidence ($B \ge 10$) against CMSSM models relative to an untuned model, except in the focus point region. Equivalently, we require at least 3 extra bits of information to fine-tune the $Z$ mass. Indeed, the connections that we have established between fine tuning, statistics and information theory allow us to understand the severity of \namebg measures using the qualitative scales in \cref{tab:scales}.

These interpretations allow us to rigorously define fine-tuning, shed fresh light on our understanding of \namebg measures, and allow a precise interpretation of hundreds of recent and historical studies of fine tuning in supersymmetric models and models of new physics.

Lastly, these insights extend far beyond weak-scale supersymmetry and fine-tuning of the weak scale. They apply anywhere fine-tuning arguments were applied, e.g., cosmology~\cite{Adams:2019kby, Diaz-Pachon:2024nsq}, dark matter~\cite{Aguirre:2004qb,Pierce:2007ut, Hertzberg:2019bvt}, axions and new particles~\cite{Peccei:1977hh,Agrawal:1997gf,Arvanitaki:2009fg,Yamada:2015waa}, the cosmological constant and inflation~\cite{Barrow:2001ks,Barnes:2017rpb}, and especially in cases of a sharp measurement that can be approximated by a Dirac function.

It is a common rationale in particle physics to introduce new symmetries and particles as a solution to the experimentally-observed fine-tuning of current theories. In many instances, however, the fine-tuning level introduced by the new symmetries and particles is omitted. Our new interpretations may allow to compare the fine-tuning level among theories precisely for the first time, and to interpret the difference within the frameworks of statistics and information-theory. Such a discussion will be performed in our upcoming work.

\begin{acknowledgments}
AF was supported by RDF-22-02-079. GH is supported by the U.S. Department of Energy under the award number DE-SC0020250 and DE-SC0020.
\end{acknowledgments}

\appendix

\section{Requirements}\label{app:requirements}

\subsection{Scale-invariant prior}

To identify the Bayes factor surface and extra information with the \namebg measure, in \cref{eq:log_prior} we assumed a scale-invariant prior for the parameter exchanged for the $Z$ mass. Priors are a thorny issue in Bayesian inference. 

The scale-invariant prior is a common choice for unknown scales, as the density is invariant under rescaling~\cite{4082152,Hartigan1964,Consonni2018}. Because the density for the logarithm is constant,
\begin{equation}
    p(\log x) = \text{const.},
\end{equation}
it is also known as a logarithmic prior. Over the whole real line, this prior would be improper and we thus considered a proper prior by truncating it between $a$ and $b$. This choice is not compatible with zero, e.g.,~$A_0 = 0$, and so our interpretation of the \namebg somewhat breaks down.

Although the scale invariant could be a reasonable choice, it is not necessarily our recommended prior. This choice of prior served to demonstrate an interpretation of the \namebg measure: \emph{if} this was our prior, we could identify the \namebg measure with a BF surface and the extra information required to fit the $Z$ mass. 

\subsection{Dirac approximation}\label{app:gaussian}

In \cref{eq:dirac_mz}, we approximated the likelihood function for the measurement of the $Z$ mass, $\hat M_Z = 91.1876\pm0.0021\gev$~\cite{Workman:2022ynf}, by a Dirac function. We now consider a Gaussian likelihood, $\hat M_Z \pm \sigma$,
\begin{equation}\label{eq:gaussian_mz}
    \like(M_Z) = \frac{1}{ \sqrt{2 \pi} \sigma} e^{-\tfrac{(M_Z-\hat{M}_Z)^2}{2\sigma^2}}.
\end{equation}
For simplicity, we denote expectations of this Gaussian as,
\begin{equation}\label{eq:notation}
    \moment{g(M_Z)} \equiv \int\limits_{-\infty}^{\infty} \like(M_Z) \, g(M_Z) \intd M_Z. 
\end{equation}
To justify our Dirac approximation, we use techniques that are related to a Laplace approximation~\cite{doi:10.1080/01621459.1986.10478240}.

\subsubsection*{Evidences}

Using our notation \cref{eq:notation}, we may the write evidences, e.g., \cref{eq:z_0,eq:z_i}, as,
\begin{equation}
\ev = \moment{\pi(M_Z \given \vtheta)},
\end{equation}
where, by the Jacobian rule, the prior density of the $Z$ mass,
\begin{equation}
\pi(M_Z \given \vtheta) = \pi(\phi \given \vtheta) \left| \frac{\text{d}\phi}{\text{d}M_Z} \right|.
\end{equation}
In this notation, the Dirac approximation leads to
\begin{equation}\label{eq:dirac_approx_z}
\ev = \pi(\hat M_Z \given \vtheta).
\end{equation}
We now compare this to results from the Gaussian likelihood in \cref{eq:gaussian_mz}. 

Assuming the prior $\pi(M_Z)$ varies slowly around the peak of the Gaussian likelihood, $\hat{M}_Z \pm \sigma$, we can perform a Taylor expansion around $\hat{M}_Z$, such that
\begin{equation}\label{eq:taylor_prior}
\begin{split}
 \pi(M_Z) = {}&\pi(\hat{M}_Z) \\
    & +  (M_Z-\hat{M}_Z)^{\phantom{1}} \, \pi^{\prime}(\hat{M}_Z) \vphantom{\frac{\pi^{\prime\prime}(\hat{M}_Z)}{2}}\\
    & +  (M_Z-\hat{M}_Z)^2 \, \frac{\pi^{\prime\prime}(\hat{M}_Z)}{2} \\
    & +  (M_Z-\hat{M}_Z)^3 \, \frac{\pi^{\prime\prime\prime}(\hat{M}_Z)}{3!} + \cdots
\end{split}
\end{equation}
The evidence thus reads
\begin{equation}
\begin{split}
\ev = {}& \pi(\hat{M}_Z) \\
        & + \moment{(M_Z-\hat{M}_Z)} \, \pi^{\prime}(\hat{M}_Z) \\
        & + \moment{(M_Z-\hat{M}_Z)^2} \, \frac{\pi^{\prime\prime}(\hat{M}_Z)}{2} \\
        & + \moment{(M_Z-\hat{M}_Z)^3} \, \frac{\pi^{\prime\prime\prime}(\hat{M}_Z)}{3!} + \cdots 
\end{split}
\end{equation}
Using the moments of a Gaussian distribution, 
\begin{subequations} \label{eq:normal_moments}
\begin{align}
    &\moment{(M_Z-\hat{M}_Z)} = 0\\
    &\moment{(M_Z-\hat{M}_Z)^2} = \sigma^2\\
    &\moment{(M_Z-\hat{M}_Z)^3} = 0 \\ 
    &\moment{(M_Z-\hat{M}_Z)^4} = 3 \sigma^4\\
    &\moment{(M_Z-\hat{M}_Z)^5} = 0
\end{align}
\end{subequations}
we obtain
\begin{equation}
    \ev = \pi(\hat{M}_Z) \left[1  + \mathcal{O}\left(\sigma^2 \frac{\pi^{\prime\prime}(\hat{M}_Z)}{\pi(\hat{M}_Z)}  \right)\right]. 
\end{equation}
Thus, the evidence equals that in the Dirac approximation, \cref{eq:dirac_approx_z}, up to second-order variations in the prior around $\hat M_Z \pm \sigma$. 

\subsubsection*{Average goodness-of-fit}

Let us now justify the Dirac approximation in the average goodness-of-fit contributions to the Bayes factor. We assumed that $\Delta\langle\ln \like\rangle$ in \cref{eq:b_decomposition} can be neglected. First, by Bayes' theorem the posterior for the $Z$ mass may be written,
\begin{equation}\label{eq:posterior_mz}
    p(M_Z \given \vtheta, \hat{M}_Z) = \frac{\like(M_Z) \, \pi(M_Z \given \vtheta)}{\ev}.
\end{equation}
Second, the average goodness-of-fit can be written as an average over the predicted $Z$ mass,
\begin{equation}
    \langle \ln \like \rangle = \int \ln \like(M_Z) \, p(M_Z \given \vtheta, \hat{M}_Z) \intd M_Z. 
\end{equation}
Using \cref{eq:posterior_mz} and our notation \cref{eq:notation}, we write this as,
\begin{equation}\label{eq:gof_average_notation}
    \langle\ln \like\rangle = \frac{\moment{\ln \like(M_Z) \, \pi(M_Z)}}{\ev}.
\end{equation}
Through \cref{eq:gaussian_mz}, we express the log-likelihood appearing in \cref{eq:gof_average_notation} as,
\begin{equation}
    \ln \like(M_Z) = -\ln\sqrt{2 \pi \sigma^2}-\frac{(M_Z-\hat{M}_Z)^2}{2 \sigma^2},
\end{equation}
such that,
\begin{equation}
\begin{split}
\langle\ln \like\rangle ={}
     -\frac{\ln\sqrt{2 \pi \sigma^2}}{\ev} \, &\moment{\pi(M_Z)}\\
     -\frac{1}{2 \sigma^2 \ev} \, &\moment{(\phi-\hat{M}_Z)^2 \, \pi(M_Z)}.
\end{split}
\end{equation}
As before in \cref{eq:taylor_prior}, we make a Taylor expansion for the prior such that
\begin{equation}
\begin{split}
\langle\ln \like\rangle =  
    -\frac{\ln\sqrt{2 \pi \sigma^2}}{\ev} \, &\moment{\pi(M_Z)} \\
    - \frac{1}{2 \sigma^2 \ev} \, &\moment{(\phi-\hat{M}_Z)^2} \, \pi(\hat M_Z)\\
    - \frac{1}{2 \sigma^2 \ev} \, &\moment{(\phi-\hat{M}_Z)^3} \, \pi^\prime(\hat M_Z) \\
    - \frac{1}{2 \sigma^2 \ev} \, &\moment{(\phi-\hat{M}_Z)^4} \, \frac{\pi^{\prime\prime}(\hat M_Z)}{2} \\
    - \frac{1}{2 \sigma^2 \ev} \, &\moment{(\phi-\hat{M}_Z)^5} \, \frac{\pi^{\prime\prime\prime}(\hat M_Z)}{3!}\\
     - \cdots\phantom{-\frac{1}{2 \sigma^2 \ev}}\negphantom{- \cdots} &
\end{split}
\end{equation}
Using the moments of a Gaussian in \cref{eq:normal_moments}, we obtain
\begin{equation}
\begin{split}
   \langle\ln \like\rangle =& -\ln\sqrt{2 \pi \sigma^2}  - \frac{\pi(\hat M_Z)}{2 \ev}\\
&- \frac{3 \sigma^2 \pi^{\prime\prime}(\hat M_Z)}{4\ev} + \cdots 
\end{split}
\end{equation}
Thus, the average goodness-of-fit may be written,
\begin{equation}\label{eq:gof}
   \langle\ln \like\rangle = -\ln\sqrt{2 \pi \sigma^2} - \frac12 + \mathcal{O}\left[\sigma^2 \frac{\pi^{\prime\prime}(\hat M_Z)}{\pi(\hat M_Z)}\right].
\end{equation}
Finally, the first two terms in \cref{eq:gof} cancel in differences in average of goodness-of-fit; thus,
\begin{equation}
    \Delta \langle\ln \like\rangle = \Delta \mathcal{O}\left[\sigma^2 \frac{\pi^{\prime\prime}(\hat M_Z)}{\pi(\hat M_Z)}\right].
\end{equation}
That is, the difference in average goodness-of-fit depends on second-order variations of the prior prediction around $\hat M_Z \pm \sigma$.

\subsection{Model priors}

To identify the posterior probability of a model with $1/\BG \times 100 \%$, we assumed that $\pi_0 = \pi_1 = 1/2$. This assumption was only necessary for \cref{eq:posterior}.

\section{Previously known connections}\label{app:history}
 
The connections between fine-tuning and Bayesian inference were previously known and discussed in e.g.~\ccite{4018651d-dd41-3908-8c4f-09d2118c5612,1991BAAS...23.1259J,mackay2003information} and explored in the specific context of weak-sale supersymmetry in \ccite{Allanach:2007qk,Cabrera:2008tj,
Cabrera:2010dh,Balazs:2012vjy,Fichet:2012sn,Fowlie:2014faa,Fowlie:2014xha,
Fowlie:2015uga,Clarke:2016jzm,Fowlie:2016jlx,Athron:2017fxj,Fundira:2017vip}. The precise connection between \namebg measures and Bayesian inference was hindered by the fact that the \namebg measure cannot be directly identified with any objects in traditional Bayesian inference.

The connections required a scale-invariant prior for the $\mu$-parameter. First, it was known that upon marginalizing parameters including the $\mu$-parameter to produce e.g., a one-dimensional posterior density, the \namebg measure for the $\mu$-parameter would appear under integration from a combination of the scale-invariant prior and a Jacobian, 
\begin{equation}\label{eq:post}
    p(x_1 \given \hat M_Z) \propto \int \frac{1}{\BG_\mu} \, p(x_1, x_2, x_3,\ldots) \intd x_2 \intd x_3 \cdots
\end{equation}
That is, the \namebg measure appears as a factor in the integrand in the posterior. The posterior, however, is a density; it depends on choice of parameterization and cannot be compared to the \namebg measure, a number. Second, it was known that the \namebg measure appeared as factor in the integrand in the Bayesian evidence for the $Z$ mass,
\begin{equation}\label{eq:z}
    \ev \propto \int \frac{1}{\BG_\mu} \, p(x_1, x_2, x_3, x_4, \ldots) \intd x_1 \intd x_2 \intd x_3 \intd x_4 \cdots
\end{equation}
The \namebg measure depends on choices of parameters; thus cannot be readily compared to the evidence, as they are marginalized in the latter. Thus, the connections \cref{eq:post,eq:z} are insightful but do not allow a direct interpretation of the \namebg measure.

\section{Minimizing fine-tuning}

The \namebg measure depends on a choice of parameters; it is common to minimize it with respect to these parameters,
\begin{equation}\label{eq:min_max}
    \min \BG \equiv \min_{\vtheta} \max_i \BG_i(\vtheta).
\end{equation}
This is a min-max equation. On the other hand, consider the Bayes factor for a model in which we marginalize every parameter according to the prior $\pi(\vtheta)$,
\begin{equation}
    B_{10} = \frac{\ev}{\ev_0} = \frac{\int \pi(\vtheta) \ev_i(\vtheta) \intd \theta}{\ev_0}
\end{equation}
By considering the maximum of the integrand, the Bayes factor must satisfy,
\begin{equation}
    B_{10} \le \frac{\max_{\vtheta} \ev_i(\vtheta)}{\ev_0}
\end{equation}
This must hold for every choice $i$. Suppose that for every parameter $i$ we picked a scale-invariant prior that spanned the same number of orders of magnitude. The ratio of evidences could be written in terms of the \namebg measure for every $i$,
\begin{equation}
    B_{01} \ge \min_{\vtheta} \BG_i(\vtheta).
\end{equation}
As this must hold for every choice $i$, the strongest bound becomes,
\begin{equation}\label{eq:bound}
    B_{01} \ge \max_i \min_{\vtheta} \BG_i (\vtheta). 
\end{equation}
We thus find that minimizing fine-tuning results in a bound on the Bayes factor. 

Our \cref{eq:bound}, however, is a max-min inequality. For every parameter, we minimize the \namebg measure, and finally take the maximum across choice of parameter. Max-min and min-max are connected by max-min inequalities~\cite{boyd2004convex},
\begin{equation}\label{eq:max_min_inequality}
\max_i \min_{\vtheta} \BG_i (\vtheta) \le \min_{\vtheta} \max_i  \BG_i (\vtheta) = \min \BG.
\end{equation}
Thus, unfortunately, we cannot chain inequalities (\ref{eq:bound}) and (\ref{eq:max_min_inequality}). Rather than computing \cref{eq:min_max}, we should consider computing \cref{eq:bound}, as it bounds the evidence against a model and the information that must be provided to tune the $Z$ mass relative to an untuned model.

\bibliographystyle{apsrev4-1}
\bibliography{refs}

\end{document}